# Luminescence Decay and Absorption Cross-Section of Individual Single-Walled Carbon Nanotubes

Stéphane Berciaud, Laurent Cognet and Brahim Lounis[*]

*Centre de Physique Moléculaire Optique et Hertzienne, Université de Bordeaux and CNRS,*

*351 cours de la Libération, Talence, F-33405 France*

[*]email: b.lounis@cpmoh.u-bordeaux1.fr

The absorption cross-section of highly luminescent individual single-walled carbon nanotubes is determined using time–resolved and cw luminescence spectroscopy. A mean value of $10^{-17}$ cm$^2$ per carbon atom is obtained for (6,5) tubes excited at their second optical transition, and corroborated by single tube photothermal absorption measurements. Biexponential luminescence decays are systematically observed, with short and long lifetimes around 45ps and 250ps. This behavior is attributed to the band edge exciton fine structure with a dark level lying a few meV below a bright one.

The solubilization of individual single walled carbon nanotubes (SWNTs) in aqueous surfactant solutions and the subsequent observation of bandgap emission from semiconducting species[1, 2], has propelled fluorescence techniques as a versatile tool for fundamental investigations of the structure related properties of these quasi ideal 1D systems. It is now well established that photon absorption in semiconducting SWNTs can lead to the formation of tightly bound excitons[3, 4]. In this context, the dynamics of optically active excitons has been studied on large ensembles of micelle-encapsulated SWNTs[5-8]. Multi-exponential decays have been reported previously[7-9], but the relative contributions of intrinsic and extrinsic effects remain a matter of debate. Since ensemble measurements are altered by extrinsic factors such as residual bundling, SWNT structure heterogeneities and length dispersions, optical studies on individual chirality-assigned SWNTs are essential. Pioneering lifetime measurements performed on single nanotubes[10] exhibited monoexponential luminescence decays with a large heterogeneity from tube to tube. The presence of numerous trap states was evoked to account for such dispersion, implying that the intrinsic properties of nanotubes were hidden. Along this line, the notion has emerged that weakly sonicated tubes observed in aqueous environments, or unprocessed air suspended tubes should be used to minimize the effects of possible surface interactions or nanotubes defects. Those two strategies agreed to infer quantum yields of individual tubes much higher[11-13] than previously estimated on strongly sonicated tubes[1,5,6,10]. Importantly, such determinations rely on the knowledge of the absorption cross-section of the nanotubes, but up to now, the large amount of data collected on the luminescence properties of semiconducting SWNTs contrasts with the lack of available information on their absorptive properties, and a quantitative measurement of the absorption cross-section of a single nanotube is still lacking.

In this letter, we report the measurement of the absorption cross-section of weakly sonicated individual HiPco nanotubes, embedded in an agarose gel. A mean value of $10^{-17}$ cm$^2$ per carbon atom is deduced from high-resolution time-resolved luminescence decays combined with cw luminescence studies on individual (6,5) SWNTs resonantly excited at their second order excitonic transition. The values of the absorption cross-section obtained are further corroborated by independent photothermal heterodyne measurements. Those results provide a good basis for a reliable estimation of the quantum yield and radiative lifetime at the single nanotube level. Interestingly, the luminescence decays of these highly luminescent nanotubes systematically exhibited a biexponential behavior. The short (30-60 ps) and long (150-330 ps) components are attributed to the excitonic fine structure of the nanotubes. The measured decay times are consistent with the existence of a dark excitonic state lying ~ 5meV below the bright exciton and suggest a slow thermalization between excitonic levels.

We prepared samples containing a fraction of highly luminescent and relatively long nanotubes by dispersing raw HiPco SWNTs in aqueous sodium dodecylbenzenesulfonate (SDBS) surfactant using very brief tip ultrasonication. The nanotubes were immobilized in a 4% agarose gel sandwiched between two glass cover slips. Samples were mounted onto an confocal microscope equipped with a high NA immersion objective (NA=1.4).

We focused our study on (6,5) nanotubes (peak emission at 975 nm)[2] which were resonantly excited at their second order excitonic resonance ($S_{22}$ at 567nm) using a cw dye laser or an optical parametric oscillator (~150fs pulses, 76MHz repetition rate). After photon absorption, the formed $S_{22}$ excitons decay on a ~40fs timescale[14] essentially into the $S_{11}$ excitonic manifold[15], from which luminescence photons are emitted. The near-IR luminescence was detected by a silicon avalanche photodiode (APD) or a 1D Nitrogen-cooled Si-CCD camera at the output of a spectrometer. Luminescence decays were recorded under pulsed excitation using a conventional time resolved single photon counting scheme. Those measurements were then followed by a study of the intensity dependent luminescence rate of the same SWNTs performed with the cw dye laser for excitation. Unless otherwise stated, the excitation beams (pulsed or cw) were linearly polarized along the axis of each individual tube. Both beams were focused onto a 550±50nm (FWHM) calibrated spot on the sample. All experiments were carried out at room temperature.

The samples were raster scanned to record 2D confocal luminescence images on which micrometer long nanotubes could be distinguished among several diffraction limited spots corresponding to shorter ones. We selected only the brightest of the optically resolved (6,5) nanotubes. Figure 1a-b show the luminescence image and narrow emission spectrum of such a bright immobilized SWNT obtained with a weak cw excitation intensity of ~1kW/cm$^2$. From 48 individual (6,5) SWNTs emission spectra, we obtained a mean peak emission value of 975±2nm and a mean homogenous full width of 20±2meV. Those narrow distributions confirm that an aqueous gel provides a minimally disturbing environment for SWNTs, where they display high and stable luminescence[16], in contrast with individual SWNTs deposited on a surface.

Figure 1c shows the luminescence decay obtained from the nanotube of figure 1a. In order to prevent multiple exciton formation, a very low photon fluence of 10$^{12}$ photons/pulse/cm$^2$ was used with 10 min integration times. We further checked that the nanotubes displayed constant emission intensities during the acquisitions (Fig. 1d) and identical spectra at the end of the acquisitions. All the tubes decays were accurately fitted by biexponential functions while monoexponential fits could not reproduce the experimental data (Fig. 1c, e-f). For the fits we used the convolution of the instrumental response function with a double exponential function $A_S \exp(-t/\tau_S) + A_L \exp(-t/\tau_L)$ with $\tau_L > \tau_S$. The statistical distributions of the short and long times built from 55 luminescence decays are shown in fig. 2a-e. The fast decay times $\tau_S$ ranged from 28 to 66 ps (mean value of 44±9 ps, Fig. 2b) while $\tau_L$ values are all found between 155 and 330 ps (mean 230±40 ps, figure 2c). No correlation is observed between the short and long decay times. In addition, the decay times are uncorrelated with the peak emission wavelengths and the weight of the fast decay, defined as $Y_S = A_S \tau_S / (A_S \tau_S + A_L \tau_L)$ (mean 80±6%).

In contrast to previous single nanotube measurements, the luminescence decays are recorded at very low excitation intensities and our weakly sonicated tubes exhibit narrow distributions of emission peaks, spectral linewidths and decay times. We therefore attribute the observed biexponential decay to the excitonic structure of defect free carbon nanotubes. We explain this behavior by the presence of a dark state in the band edge exciton fine structure lying a few meV below a bright one. The existence of such states has been predicted theoretically[9, 17-19], and recently studied experimentally[20-22].

In our model, we consider a three level system (Fig. 2f) consisting of the zero exciton ground level $|G\rangle$, and two excitonic levels $|D\rangle$ and $|B\rangle$, representing respectively, the dark, (even parity), and the bright (odd parity) singlet excitons, with recombination rates $\Gamma_B$ and $\Gamma_D$. Thermalization between dark and bright states can occur through coupling to acoustic phonon modes whose energies matches the dark-bright excitonic splitting $\Delta E \sim 5$ meV [20, 22]. The transition rates for downhill and uphill processes write $\gamma_\downarrow = \gamma_0 (n+1)$ and $\gamma_\uparrow = \gamma_0 n$, respectively, where $n(T) = 1/[\exp(\Delta E / k_B T) - 1]$ is the Bose-Einstein phonon number at temperature T and $\gamma_0$ the zero temperature bright to dark transition rate. By solving the kinetic equations, we derive the dynamics of the bright excitonic state population $P_B(t)$ after excitation at t=0. Neglecting radiative recombination from the dark state, the luminescence signal is simply proportional to $P_B(t)$ and shows a biexponential behavior. Assuming $1/\Gamma_B$ = 60ps, $1/\Gamma_D$ = 600ps, $1/\gamma_0$ = 1300ps and $P_B(0) = 1$ we obtain short and long time constants of $\tau_S$ = 45ps and $\tau_L$ = 215ps and a fractional yield $Y_S$ = 78% which reproduces quantitatively the experimental decay of figure 1c.

Interestingly, we found that a relatively low value of $\gamma_0$ ($\gamma_0 \ll \Gamma_B$) is needed to account for our experimental finding. This suggests that symmetry breaking processes caused by external perturbations[18] are sufficiently weak, so that exciton thermalization does not occur prior to emission.

We now consider the determination of the absorption cross-section of an individual nanotube. In a simple model where multi-excitonic processes do not play a role, an analysis of the deviation from linearity of the luminescence signal as a function of cw excitation intensity is sufficient to retrieve the absorption cross-section of an individual nanotube provided its exciton decay time is also known. Therefore, we define the *effective lifetime* $\tau_X = \Gamma_X^{-1}$ of the exiton as $\tau_X = \left(\dfrac{Y_S}{\tau_S} + \dfrac{1-Y_S}{\tau_L}\right)^{-1}$. Using the statistical distributions of figure 2, we find a mean value of $\langle \tau_X \rangle = 53 \pm 10 \,\mathrm{ps}$.

We introduce $N_{abs}(L) = \sigma_{22} L \dfrac{I}{\hbar\omega}$, the rate of photons absorbed by a nanotube segment of arbitrary length $L$ interacting with a laser having an intensity $I$, a polarization parallel to the tube axis and a photon energy $\hbar\omega \sim 2.19\,\mathrm{eV}$ resonant with the second order transition of (6,5) nanotubes. In this expression, $\sigma_{22}$ is defined as the resonant absorption cross-section of the nanotube per unit length. In the following, we will assume a diffusional excitonic motion characterized by an exciton diffusion range $\Lambda \sim 90\,\mathrm{nm}$ [16], which is the average nanotube length explored by an exciton during its lifetime. In the regime where $N_{abs}(\Lambda) \leq \Gamma_X$, the luminescence signal is exclusively due to monoexcitonic radiative recombination. The detected counting rate thus writes $N(L) = N_0 \dfrac{N_{abs}(L)/\Gamma_X}{1 + N_{abs}(L)/\Gamma_X}$, where $N_0$ is the saturated counting rate due to the finite exciton decay rate.

Figure 3a shows the detected counting rate of an individual (6,5) nanotube as a function of cw laser intensity. Given the effective lifetime of this nanotube ($\tau_X = 55 \pm 5\,\text{ps}$), previously deduced from its time resolved luminescence decay (see inset), and $L = 550 \pm 50\,nm$ (the size of the excitation beam), a fit of the experimental counting rate with the saturation profile described above yields a resonant absorption cross-section of $\sigma_{22}$ = 110±15 nm²/µm. Using this measurement, we find that at the maximum cw excitation intensity (~15kW/cm²) used in figure 3a, the absorption rate per exciton diffusion range $\Lambda$ ($N_{abs}(\Lambda) \sim 4\text{ns}^{-1}$) is significantly smaller than $\Gamma_X = 18\,\text{ns}^{-1}$. In these conditions there is always less than one exciton created (~0.2) per lifetime and per exciton diffusion range in the nanotube, which validates the monoexcitonic assumption used to deduce $\sigma_{22}$.

The histogram of the absorption cross-sections of 27 individual SWNTs is presented in figure 3b and shows a distribution with a mean value of $\sigma_{22}$ ~ 85 nm²/µm and a dispersion of ± 30 nm²/µm around this value. This translates into a mean absorption cross-section of (1 ± 0.3) x $10^{-17}$cm² per Carbon atom, a value one order of magnitude larger than the commonly used value of ~$10^{-18}$ cm²/C atom[23]. The latter has been obtained from ensemble absorption spectra, where the inhomogeneously broadened feature assigned to the $S_{22}$ transition may stem from several SWNT structures and from nanotubes possibly aggregated into small bundles..

To confirm our values of $\sigma_{22}$, we measured the absorption of individual nanotubes using the photothermal heterodyne method[24, 25]. For this purpose, we compared the amplitudes of the photothermal heterodyne signal arising from individual (6,5) carbon nanotubes (Fig. 3d) with that of individual spherical gold nanoparticles (mean diameter of 9.2±1nm) with known absorption cross-section (Fig. 3e). The nanotubes and the nanoparticles were both excited with circularly polarized light at the same excitation wavelengths (567 nm) and laser intensities. The photothermal effect (absorption-induced local refractive index changes) was probed by means of a non-resonant Nd:Yag laser (1064nm). Knowing the absorption cross-section of the gold nanoparticles at this wavelength (~25 nm$^2$ [26]) and taking into account the laser polarization and the point spread function of the apparatus (550nm), we deduce $\sigma_{22}$ ~ 90nm$^2$/μm for the (6,5) nanotubes, corroborating the values estimated from luminescence measurements.

Finally, we estimate the quantum yield $\eta$ of individual nanotubes, using the expression $\eta = \dfrac{N_0}{\eta_D \Gamma_X}$, where $\eta_D$ ~ 10$^{-3}$ is the overall detection efficiency of our experimental setup. The estimated quantum yields range from ~0.1% to ~1% (Fig. 3e) which compares well with recent estimations[11-13, 27]. We can also estimate a mean value of the radiative lifetime of ~12±4 ns, in good agreement with recent theoretical predictions[18, 19] as well as estimations from ensemble measurements[6]. Interestingly, figure 3c reveals a clear correlation between the effective lifetime of a single tube and its quantum yield, the brightest tubes having the longest lifetimes. The lowest quantum yield values probably result from the presence of defects created during the brief sonication process. Such defects generate trap states, which open new non-radiative decay channels, resulting in shorter excitonic lifetimes and thus lower quantum yields.

In conclusion, we have performed a detailed room temperature study of the luminescence dynamics of individual (6,5) nanotubes. All the luminescence decays exhibited a biexponential behavior which highlights the coupling between dark and bright excitonic states. Using cw excitation on the same nanotubes, we deduced resonant absorption cross-sections that was independently confirmed by photothermal absorption measurements. This work will be essential in proper estimation of excitonic population in further nanotube characterizations including temperature dependence of the luminescence decay and studies of multiexcitonic effects.

Acknowledgements:

We warmly thank D. A. Tsyboulski and R. B. Weisman for valuable advices in sample preparation. We acknowledge W. Benharbone and J.-N. Destouet for their technical support. This research was funded by Région Aquitaine, Agence Nationale de la Recherche (PNANO program), CNRS and the French Ministry for Education and Research (MENRT).

**Figure Captions:**

**Figure 1 (color online)**

(a) Confocal luminescence image of an individual (6,5) nanotube (scale bar is 1µm). (b) Luminescence spectra recorded before (line) and after (circles) the luminescence decay. (c) Time resolved luminescence decay from the nanotube shown in (a) and (b). The gray curve represents the instrumental response function. The data were recorded at very low fluence ~$10^{12}$ photons/pulse/cm$^2$, yielding an average generation of 0.08 exciton per femtosecond pulse and per exciton diffusion range. The decay is fitted by a single (blue line, residuals in (e)) and double (red line, residuals in (f)) exponential function. From the biexponential fit we deduce short and long time constants of 46ps and 210ps, respectively, as well as a short component fractional yield of 78%. (d) Time trace of the luminescence signal recorded while measuring the decay.

**Figure 2 (color online)**

(a) Long decay times as function of short ones, deduced from 48 luminescence decays of individual (6,5) swnts. (b) and (c) are the corresponding histograms. Weights of the short decay components (d) and peak emission wavelengths (e) as a function of the short decay times. (f) Three level scheme used in the model.

**Figure 3 (color online)**

(a) Luminescence signal of an individual (6,5) nanotube as a function of cw laser intensity (circles: experimental data; solid line: fit using the saturation profile described in the text). The corresponding luminescence decay (circles) and a biexponential fit (red line) are shown in the inset. Knowing the effective decay time of 55ps, we deduce a resonant absorption cross-section of 110nm$^2$/µm. (b) Histogram of the absorption cross-sections deduced for 27 individual (6,5) SWNTs. Histogram of the photothermal signals measured in the same experimental conditions on 9 individual (6,5) SWNTs (c) and 140 individual gold nanoparticles with mean diameter of 9.2nm (d). The Gaussian lines in (b-d) are guides to the eye. (e) Plot of the estimated quantum yields as a function of the effective lifetime.

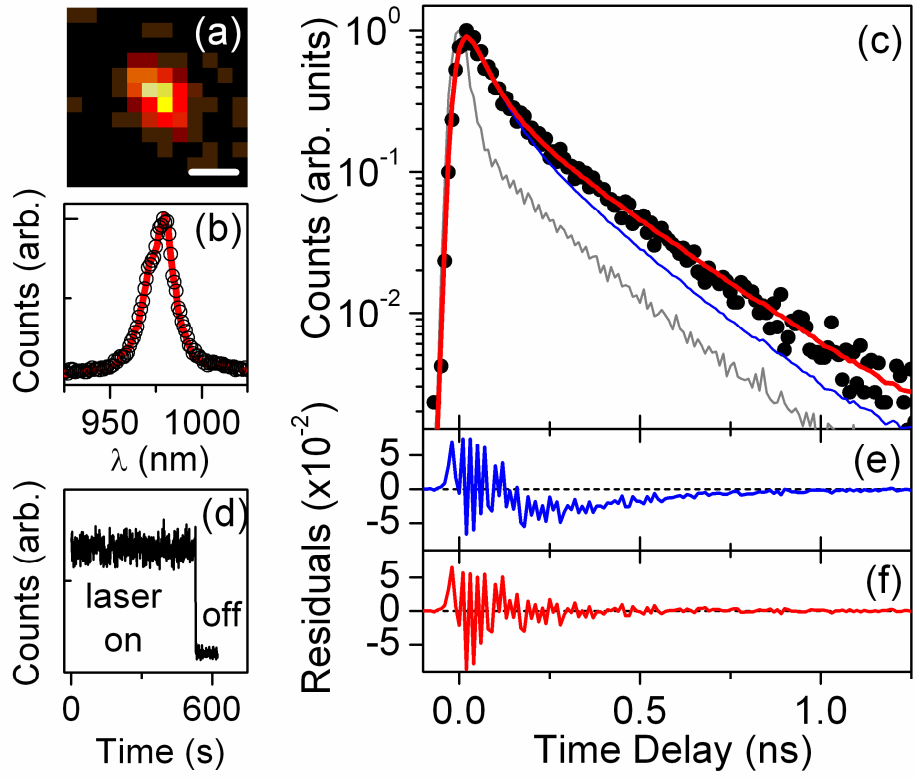

Figure 1

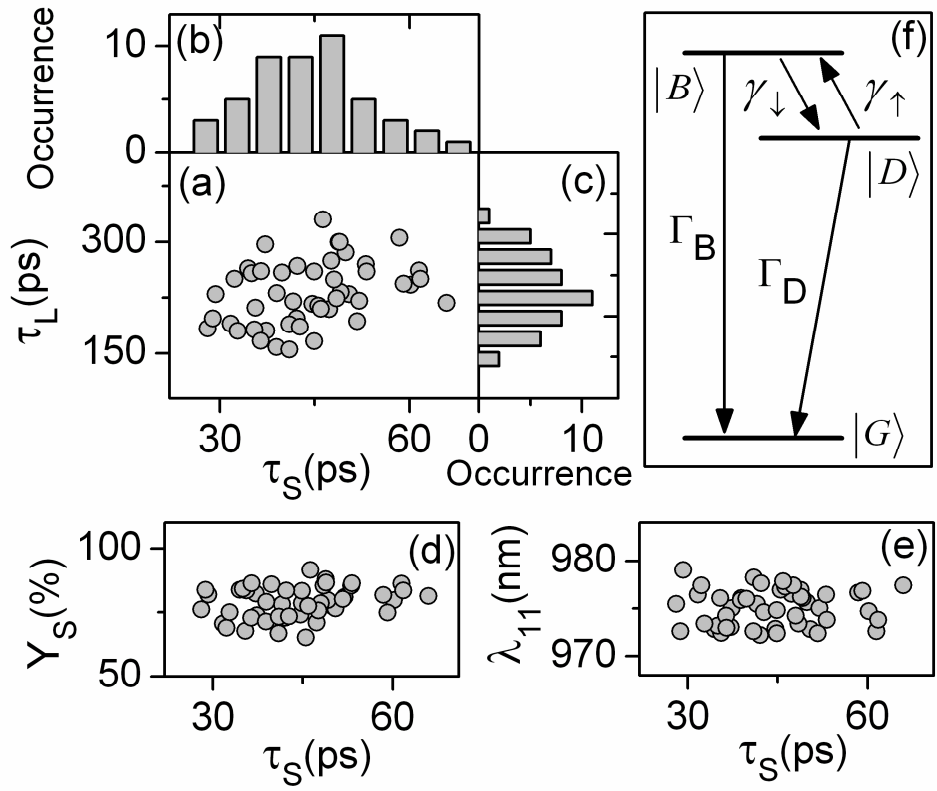

Figure 2

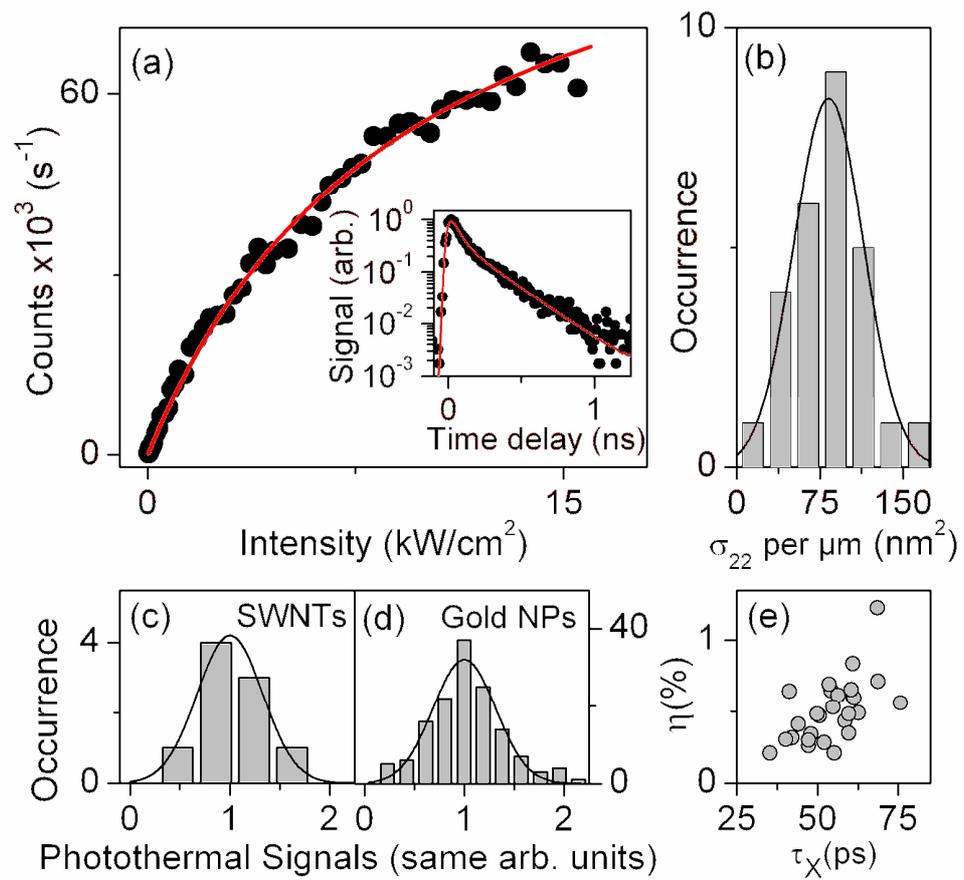

**Figure 3**